\input{aipcheck}

\documentclass[
    ,final            
  ]
  {aipproc}

\layoutstyle{6x9}

\def\be{\begin{equation}}
\def\ee{\end{equation}}
\def\baray{\begin{eqnarray}}
\def\earay{\end{eqnarray}}

\begin{document}

\title{Brane Decay and Defect Formation}


\author{Horace Stoica}{
  address={Theoretical Physics, 
  Blackett Laboratory, 
  Imperial College, 
  London SW7 2AZ U.K.}
  }

\begin{abstract}
Topological defects are generically expected to form in models of brane
inflation. Brane-anti-brane annihilation provides a way to gracefully end
inflation, and the dynamics of the tachyon field results in defect formation.
The formation of defects has been studied mainly from the brane world-volume point
of view, but the defects are themselves lower-dimensional branes, and as a
result they couple to bulk fields. 
To investigate the impact of bulk fields on brane defect formation,
we construct a toy model that captures the essential features of the
tachyon condensation with bulk fields.  In this toy model, we study the structure of
defects and simulate their formation and evolution on
a lattice.
We find that while bulk fields do not have a significant effect on defect formation, 
they drastically influence the subsequent evolution
of the defects, as they re-introduce long-range interactions between them. 
\end{abstract}

\maketitle

\section{Introduction}

Among the successes of the brane world model  is the realization of the
inflationary universe \cite{Dvali:1998pa}. The brane-anti-brane separation
plays the role of the inflaton and the non-zero attractive interaction
generates the inflaton potential. Moduli stabilisation has been recently
added to the brane-world model \cite{Kachru:2003sx}, leading to even more 
realistic brane inflation models. 

The brane inflation model contains a natural mechanism for ending inflation
via the brane-anti-brane annihilation process \cite{Sen:1998sm}. When the
inter-brane separation decreases below a critical value, the tachyon field 
which corresponds to the open string stretching between the brane and the
anti-brane develops an instability (Fig. \ref{Tachyon_Field}), and the 
rolling of the tachyon field
signals the decay of the brane-anti-brane pair. The tachyon field for a 
brane-anti-brane pair is a complex field and it has a non-trivial vacuum 
manifold which leads to the formation of topologically stable vortex
configurations. These vortices are themselves lower-dimensional branes  
\cite{Kraus:2000nj,Takayanagi:2000rz} and their formation 
\cite{Jones:2003da} and 
stability \cite{Copeland:2003bj} have been studied for a variety of 
brane inflation models. These branes couple with bulk fields and would
appear as cosmic strings to a 4-dimensional observer, therefore they are 
expected to provide a direct observational window into String Theory.

\subsection{Brane-Anti-Brane Annihilation}
\begin{figure}[t]
  \includegraphics[height=.16\textheight]{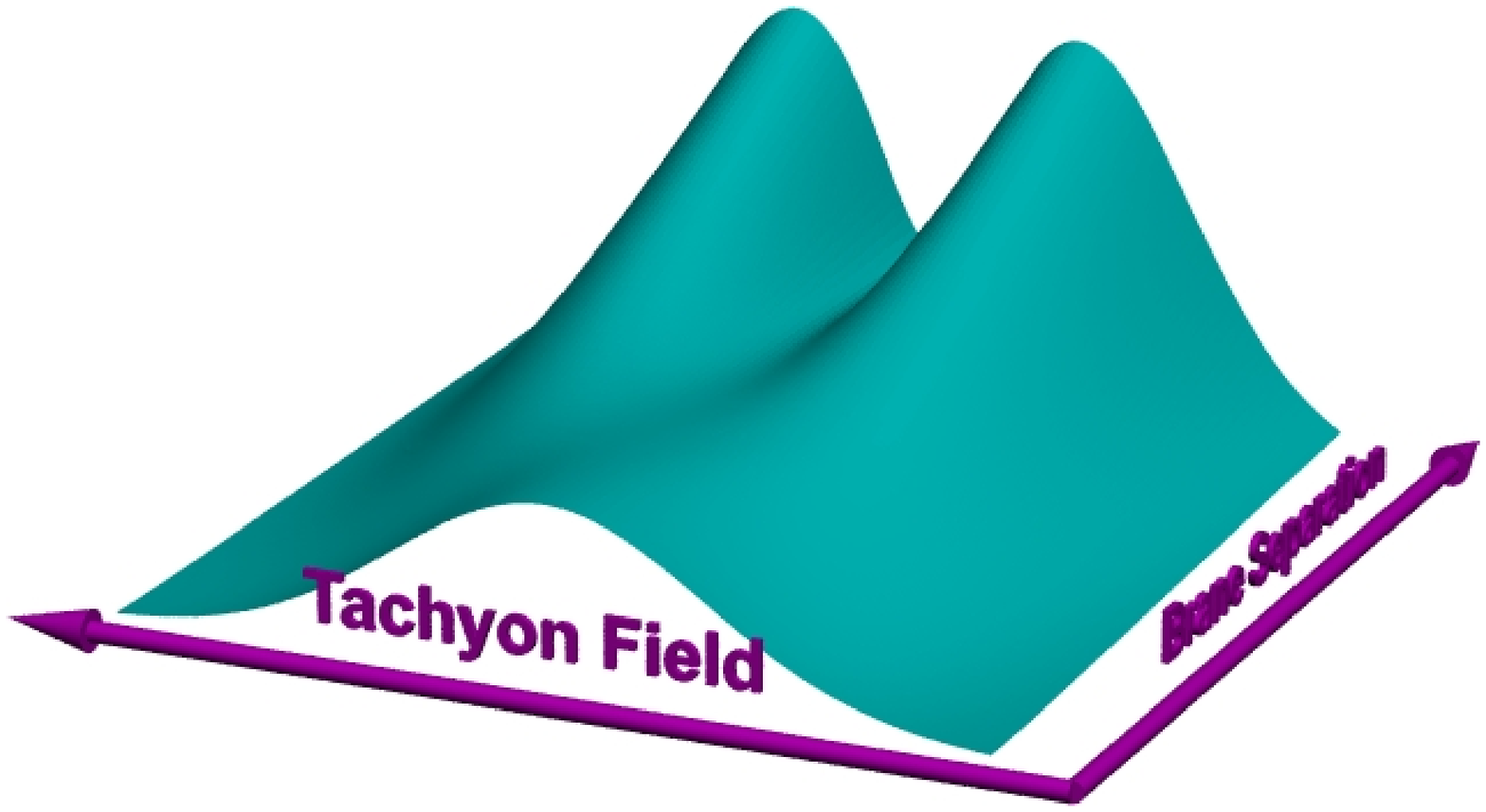}
  \hspace{6pt} 
  \includegraphics[height=.16\textheight]{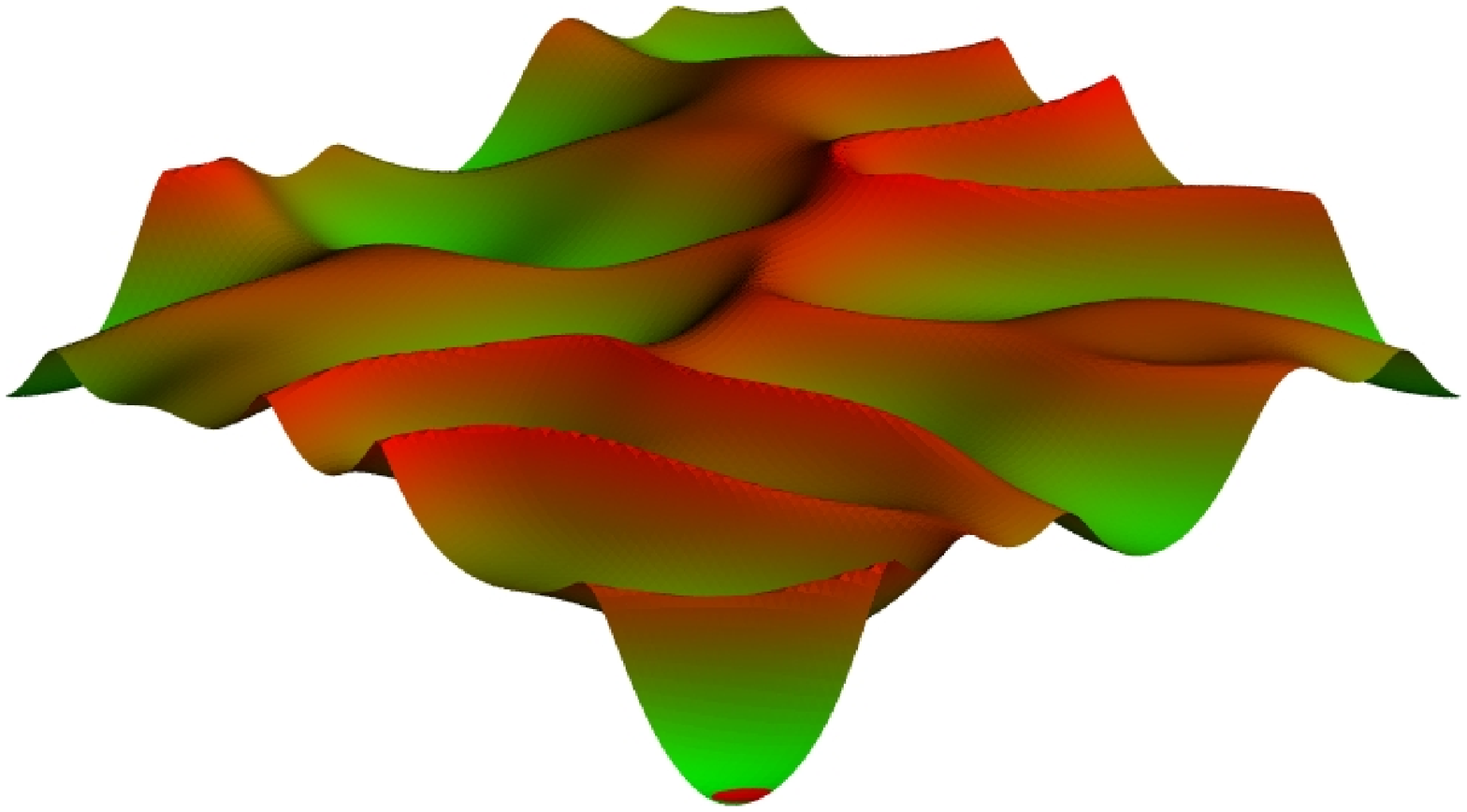}
  \caption{\label{Tachyon_Field} Left: The potential of the tachyon field as a
  function of the inter-brane separation. Right: The configuration of the
  rolling tachyon field during the brane-anti-brane annihilation process. }
\end{figure}

The action for the tachyon field of a coincident brane-anti-brane pair 
has been computed in \cite{Kraus:2000nj,Takayanagi:2000rz}. 
Neglecting the gauge fields present in the brane and the anti-brane, 
the action has the form:
\be
  S=\int d^{p+1}x\sqrt{-g}e^{-T\overline{T}/2M_{s}^{2}}
  \left(1+\partial_{\mu}T\partial^{\mu}\overline{T}\right)
\ee
Performing a lattice evolution of the tachyon field \cite{Barnaby:2004dz}, 
one notices that the field develops singularities 
[Fig. \ref{Tachyon_Field}, right panel].  These singularities are 
the locations of the
lower-dimensional branes formed in the brane annihilation process. In
\cite{Kraus:2000nj,Takayanagi:2000rz} it was shown that a linear 
spatial tachyon profile reproduces the correct tension of the lower 
dimensional branes in the limit of
infinite slope, so the occurrence of singularities was expected. 
One can use the lattice evolution of the tachyon field to estimate 
the density of vortices formed in brane annihilation, and the results
suggest a much larger density than the one estimated via the Kibble
mechanism \cite{Barnaby:2004dz}. However, the occurrence of singularities 
does not allow one to study the evolution of the network of vortices. 

\subsection{Coupling to Bulk fields}

To obtain a more complete picture of the tachyon vortex formation 
we have to include the effects of all the fields involved. 
The tachyon field is charged under a linear combination of brane 
world-volume gauge fields and this same linear combination couples 
to a bulk antisymmetric tensor fields via a Chern-Simons (CS) term. 
The relevant terms in the action are \cite{Kraus:2000nj,Takayanagi:2000rz}:
\baray
    && S=\frac{1}{2\kappa_{10}^2}\int \sqrt{-G} \: d^{10}x \left[
      e^{-2\phi}\left(R+2\nabla\phi^2\right)-\frac{F_{p}^2}{2p!}\right]
      + 2\pi \alpha^{\prime}T_{Dp}
    \int C_{p-1}\wedge \left(F^{+}-F^{-}\right)
     \nonumber \\
    &&-2T_{Dp}\int d^{p+1}x \,e^{-\phi}\, 
    e^{-2\pi\alpha^{\prime}T\overline{T}}\biggl[
      1+8\pi\alpha^{\prime}\ln\left(2\right)D^{\mu}\overline{T}D_{\mu}T+
      \frac{\gamma{\alpha^{\prime}}^{2}}{8}
      \left(F^{+}_{\mu\nu}-F^{-}_{\mu\nu}\right)^2\biggr]
      \nonumber
\earay

\begin{figure}[b]
  \includegraphics[height=.16\textheight]{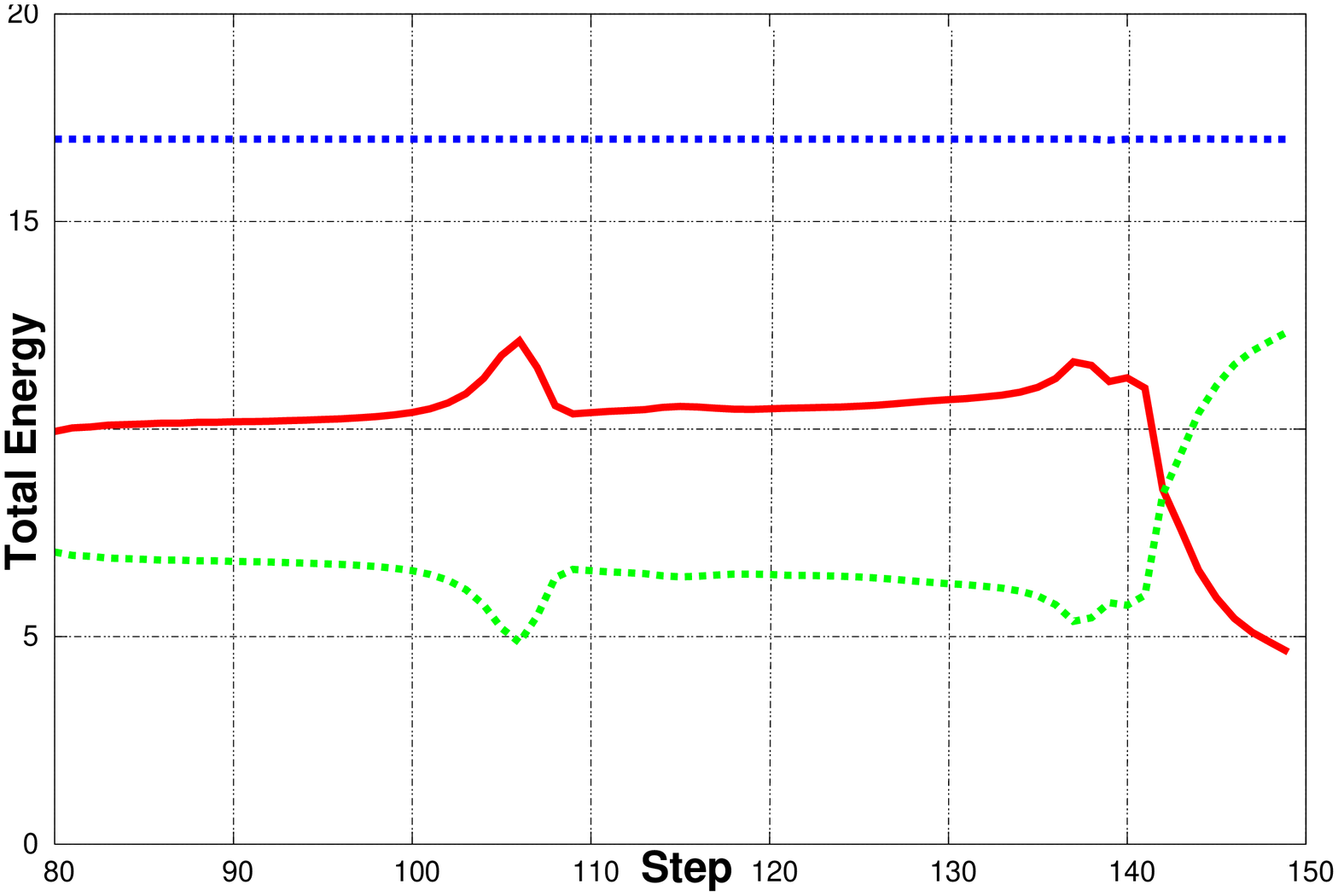}
  \hspace{6pt}
  \includegraphics[height=.16\textheight]{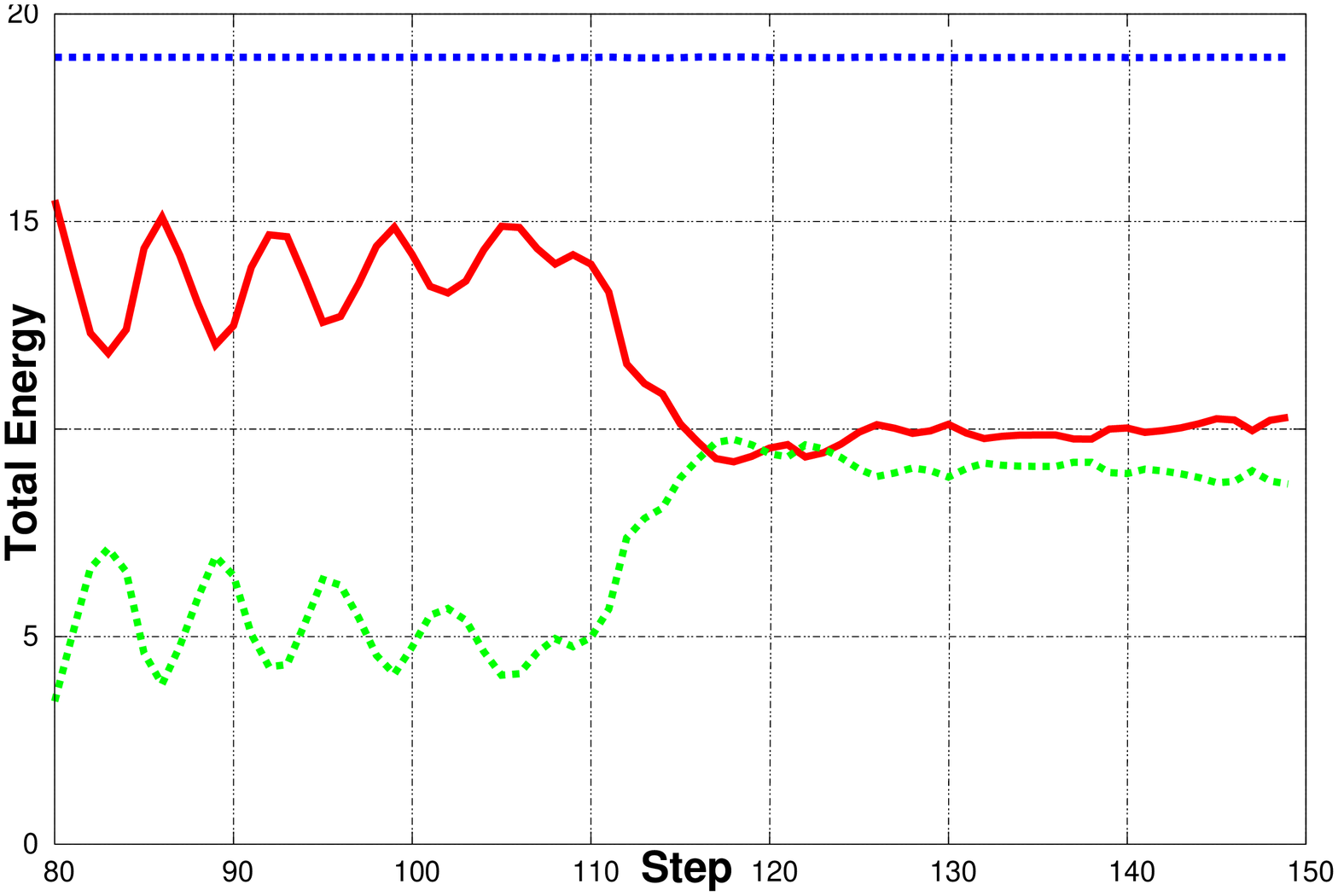}
  \caption{Effect of the size of the extra dimension on the energy of a defect
  pair. Left panel: the extra dimension is large and the bulk is effectively 
  3-dimensional. Right panel: the extra dimension is comparable to the size of 
  the defect core, making the model effectively 2-dimensional. This also
  changes the strength of the interaction between the vortices, which annihilate
  faster when the extra dimension is small.\label{Energy}}
\end{figure}

\begin{figure}[t]
  \includegraphics[height=.18\textheight]{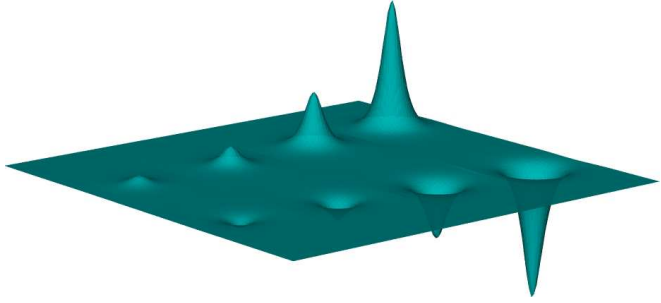}
  \caption{The magnetic flux at the core of the defect as a function of the
  CS coupling. The larger the coupling, the smaller the flux. \label{Vortex_Mag_Field}}
\end{figure}

We want to study the formation and evolution of the tachyon vortices 
using the full non-linear equations of motion. Since the above action is 
10-dimensional, we will first build a lower-dimensional toy model that 
captures the main features of the full 10-dimensional model and is amenable 
to study via a lattice simulation. We study the formation of vortices, so the 
minimal dimensionality of a brane that allows vortices to form is 2. We choose
the bulk to also have the minimum possible dimensionality, 3,
and replace the tachyon with a Higgs field in order to avoid
singularities. The rest of the correspondence between the full and the toy
model is presented in Table \ref{Toy_Model_Tab}.
\begin{table}
\begin{tabular}{cccc}
\hline
  & \tablehead{1}{c}{b}{10 D Model}
  & \tablehead{1}{c}{b}{Toy Model}   \\
\hline
   brane dim. & $p+1$   & $2+1$ \\
   bulk dim. & $9+1$   & $3+1$ \\
   scalar field & tachyon, $\; T$   & Higgs, $\; \phi$ \\
   vector field & $A_{\mu}^{+}-A_{\mu}^{-}$   & $A_{\mu}$ \\
   tensor field & $C_{M_{1}\dots M_{p-1}}$   & $C_{M}$ \\
   gravity & $G_{MN}$   & flat space \\
   coupling strength & dilaton, $\; \phi$   & fixed coupling \\
   potential & $e^{-\left|T\right|^2}$   & 
    $\lambda\left(\left|\phi\right|^2-\eta^2/2\right)^2 $\\
\hline
\end{tabular}
\caption{The correspondence between the full 10-dimensional 
model and the toy model}
\label{Toy_Model_Tab}
\end{table}
The action for the toy model follows that of the 10-dimensional model:
\be
  {\mathcal L} = -\int_{{\mathcal M}_{3}} \!\!\!dx^2dt 
    \left[\frac{F^2}{4g_{\rm brane}^2} +
      D_{\mu}\phi D^{\mu}\phi^{*}+V\left(\phi\right)\right] 
  -\frac{c_{cs}}{2}\int_{{\mathcal M}_{3}} \!\!\!F \wedge C -
  \int_{{\mathcal M}_{4}} \!\!\!dx^3dt \frac{H^2}{4g_{\rm bulk}^2}
    \nonumber 
\ee
The details of the lattice regularisation of the above action are presented 
in \cite{Moore:2006ec}. The important results are presented in Fig. 
\ref{Energy} and \ref{Vortex_Mag_Field}. Reducing the size of the extra
dimension forces the bulk field to have a larger gradient along the extra 
dimension, therefore increasing the total energy of the defect-anti-defect 
pair. The most dramatic effect is that of changing the strength of the
CS coupling. Increasing the coupling reduces the quantum of brane gauge 
field magnetic flux trapped at the core of the defect, at large values of the
coupling the flux being almost completely eliminated. This can be understood 
by first performing a dimensional reduction to 2 dimensions and then writing
the equation of motion for the bulk field. The  bulk field is 
sourced by the magnetic flux of the brane gauge field and this magnetic flux 
is localised at the defect core. 
\be
\nabla^2 C^0 = c_{\rm cs} g^2_{\rm bulk}F_{12}/2\pi R
\ee
We can use the solution for $C^0$ to find the energy density of the fields
away from the defect core.
  \be
  \nonumber 
  {\mathcal E} = | D_\theta \phi |^2 
  + \frac{2\pi R}{2 g_{\rm bulk}^2}|\nabla C^0|^2
  \ee
where $D_\theta \phi = (i n/\rho - i A_\theta) \phi$, 
$\phi^*\phi = v^2/2$, $A_\theta = \Phi/2\pi\rho$ and $C^0(\rho)  =
\frac{c_{\rm cs} g^2_{\rm bulk}}{2\pi R} \frac{\Phi}{2\pi} \ln(\rho/\rho_0)$.
$\Phi$ is the brane gauge field magnetic flux. Minimising the energy
density with respect to $\Phi$ gives the results:
\be
  \nonumber
  {\mathcal E} = \frac{v^2 n^2}{2\rho^2} \frac{a}{1+a} \, , \qquad 
    \Phi  =  \frac{2\pi n}{1 + a}\, , \qquad
  a \equiv \frac{c_{\rm cs}^2 g_{\rm bulk}^2}{2\pi R v^2} \, .
\ee
 As the CS coupling is increased the magnetic flux unit at the defect core
  decreases and the energy of the defect interpolates smoothly between 
  that of a  {\em local} and that of a {\em global} defect. 
  The bulk field also re-introduces
  long-range interactions between the local vortices.

\begin{theacknowledgments}
I would like to thank N.~Barnaby, A.~Berndsen, J.~M.~Cline and G.~D.~Moore for
collaboration on the work presented here.
\end{theacknowledgments}

\bibliographystyle{aipprocl} 

\IfFileExists{\jobname.bbl}{}
 {\typeout{}
  \typeout{******************************************}
  \typeout{** Please run "bibtex \jobname" to optain}
  \typeout{** the bibliography and then re-run LaTeX}
  \typeout{** twice to fix the references!}
  \typeout{******************************************}
  \typeout{}
 }

\end{document}